\documentclass[journal]{IEEEtran}
\usepackage{textcomp}
\usepackage{xcolor}
\usepackage{graphicx}
\usepackage{amsmath}
\usepackage{amssymb}
\usepackage{amsfonts}
\usepackage{epsfig}
\usepackage{epstopdf}
\usepackage{hhline}
\usepackage{stfloats}
\usepackage{cite}
\usepackage{algorithm}
\usepackage{algorithmic}
\usepackage[CJKbookmarks=true,
bookmarksnumbered=true,
bookmarksopen=true]{hyperref}
\usepackage{multirow}
\usepackage{subfigure}

\begin{document}
\title{Exploring the Potential of Large Language Models for Massive MIMO CSI Feedback}

\author{{\IEEEauthorblockN{Yiming~Cui\IEEEauthorrefmark{1},
 Jiajia~Guo\IEEEauthorrefmark{1},
 Chao-Kai~Wen\IEEEauthorrefmark{2},
      Shi~Jin\IEEEauthorrefmark{1},
      En~Tong\IEEEauthorrefmark{3}\ }
      
  \IEEEauthorblockA{\IEEEauthorrefmark{1}National Mobile Communications Research Laboratory, Southeast University, Nanjing 210096, P. R. China\\Email:
  cuiyiming@seu.edu.cn,
  jiajiaguo@seu.edu.cn,  jinshi@seu.edu.cn}
  
  \IEEEauthorblockA{\IEEEauthorrefmark{2}Institute of Communications Engineering, National Sun Yat-sen University, Kaohsiung 80424, Taiwan \\Email: chaokai.wen@mail.nsysu.edu.tw}

  \IEEEauthorblockA{\IEEEauthorrefmark{3}Research and Development Center, China Mobile Group Jiangsu Co., Ltd, Nanjing 210029, China \\Email: tonge@js.chinamobile.com}

 }

}	

\maketitle

\begin{abstract}

Large language models (LLMs) have achieved remarkable success across a wide range of tasks, particularly in natural language processing and computer vision. This success naturally raises an intriguing yet unexplored question: Can LLMs be harnessed to tackle channel state information (CSI) compression and feedback in massive multiple-input multiple-output (MIMO) systems? Efficient CSI feedback is a critical challenge in next-generation wireless communication. In this paper, we pioneer the use of LLMs for CSI compression, introducing a novel framework that leverages the powerful denoising capabilities of LLMs---capable of error correction in language tasks---to enhance CSI reconstruction performance. To effectively adapt LLMs to CSI data, we design customized pre-processing, embedding, and post-processing modules tailored to the unique characteristics of wireless signals. Extensive numerical results demonstrate the promising potential of LLMs in CSI feedback, opening up possibilities for this research direction.

\end{abstract}

\section{Introduction}

Massive multiple-input multiple-output (MIMO) is widely recognized as one of the key enabling technologies in fifth-generation (5G) cellular communications and is expected to play a vital role in sixth-generation (6G) cellular communications \cite{wang2024tutorial}. By equipping the base station (BS) with tens or hundreds of antennas, spectral and energy efficiency can be significantly improved \cite{marzetta2015massive}. To fully exploit the potential of massive MIMO, the BS requires accurate channel state information (CSI) \cite{lu2014overview}. The user equipment (UE) is required to feed back the estimated downlink CSI to the BS, leading to overwhelming communication overhead due to the large number of antennas in massive MIMO systems.
 
To alleviate uplink resource consumption caused by CSI feedback, extensive research efforts have been made with three mainstream methodologies: codebook-based feedback \cite{love2008overview}, compressive sensing-based feedback \cite{metzler2016denoising}, and deep learning (DL)-based feedback \cite{wen2018deep}. Among different CSI feedback methods, DL-based approaches stand out due to their impressive feedback performance and high computational efficiency. Extensive studies have been conducted to further enhance the performance and practicality of DL-based CSI feedback \cite{guo2022overview}, mainly focusing on specific neural architecture design \cite{yu2020ds}, expert knowledge-assisted design \cite{wang2021compressive,peng2024deep}, and more. However, the deployment of DL-based CSI feedback still faces severe challenges. Limited by the availability of training data and model capacity, existing DL-based methods underperform in complex and dynamic scenarios.

First, collecting field data through channel measurements is expensive, requiring complex engineering efforts and high hardware costs. Although simulated synthetic datasets can be used for training, the distribution shift between field datasets and synthetic datasets leads to performance degradation. Second, existing research primarily adopts small-scale neural networks for CSI feedback, limiting their capability to extract complex channel features and generalize across different scenarios. While training a large-scale neural network dedicated to CSI feedback could be a potential solution, it would result in high computational costs.

Transfer learning is widely recognized as a cost-effective solution when training data and computational resources are limited. By leveraging knowledge from source tasks to assist target tasks, training costs can be significantly reduced by avoiding redundant learning. The effectiveness of transfer learning heavily depends on the pre-trained model from source tasks. In recent years, the success of pre-trained large language models (LLMs) has revolutionized various industries, such as GPT \cite{radford2019language}. By training large-scale Transformer-based neural networks with massive text data, LLMs demonstrate superior and comprehensive performance across different downstream tasks. Due to their next-word prediction training strategy, LLMs also exhibit outstanding performance in time series forecasting tasks \cite{nie2022time}. This feature has been exploited in wireless communications, including channel prediction \cite{liu2024llm4cp}.

\begin{figure*}[t]
    \centering
    \includegraphics[width=0.95\textwidth]{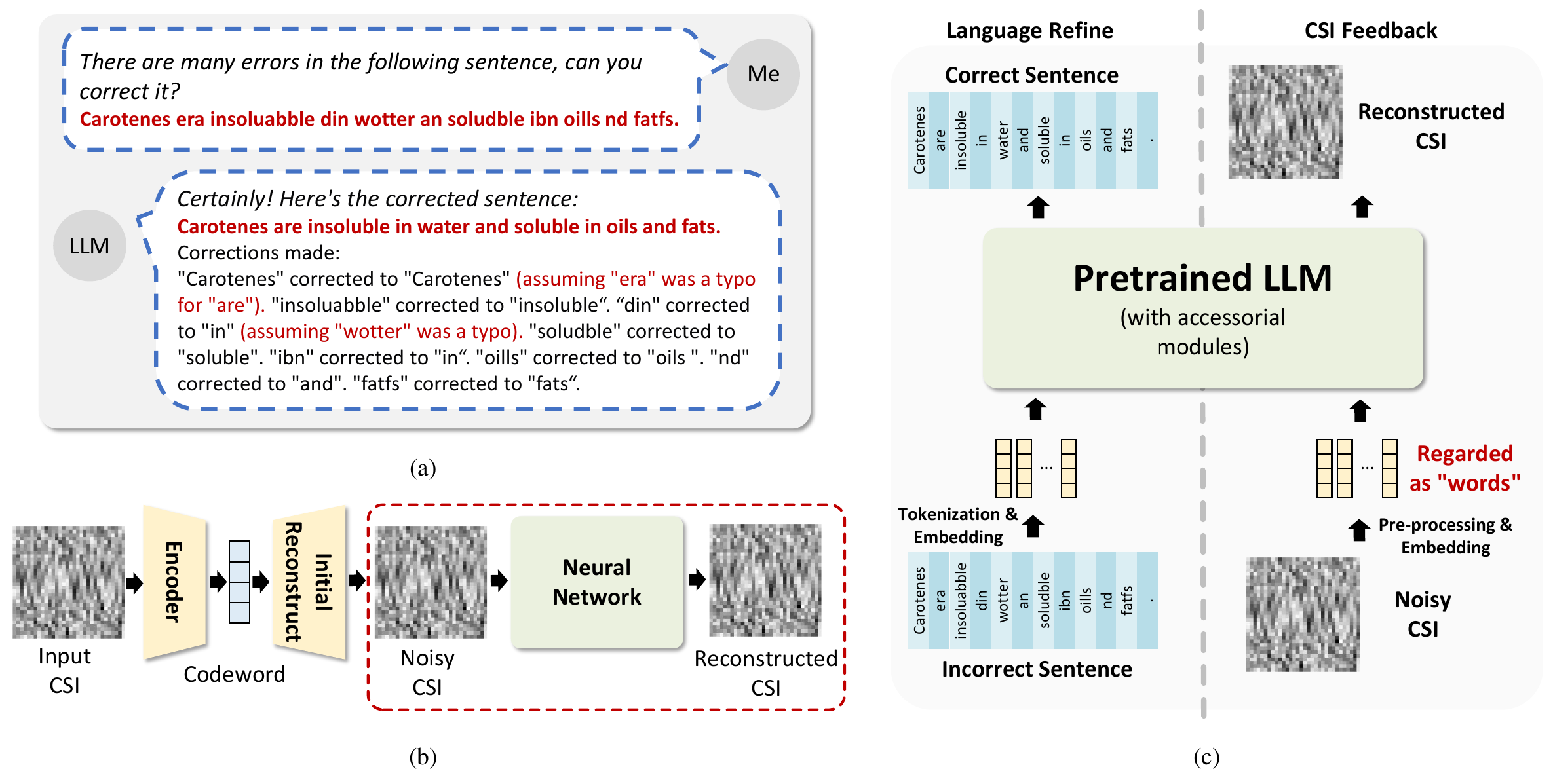}
        \caption{(a) An example of LLM-based sentence correction. (b) One-sided DL-based CSI feedback, primarily utilizing a neural network at the BS for CSI reconstruction. (c) Comparison between LLM-based sentence correction and CSI reconstruction.} 
	\label{fig:comp}
\vspace{-0.5cm}
\end{figure*}

Besides predictive tasks, one of the common applications of LLMs is language polishing. LLMs are often used to correct various errors in input sentences. In Fig. \ref{fig:comp}(a), we provide an example of error correction using an LLM. By inputting a sentence with errors, the LLM successfully outputs the corrected sentence. Two key observations can be made from this example: 
\begin{itemize}
    \item \textbf{Word Correction}: Misspelled words can be accurately corrected by the LLM. For example, the word ``insoluabble'' is corrected to ``insoluble,'' and ``fatfs'' is corrected to `fats.'' Since word errors can be regarded as noise in word tokens, this example highlights the denoising potential of LLMs. 
    
    \item \textbf{Correlation Extraction}: When processing ambiguous words, the LLM can utilize in-context information for error correction. For example, the word ``in'' is mistaken as ``din'' in the sentence, which could also be misinterpreted as ``dine'' if viewed in isolation. However, the LLM leverages the subsequent word ``water'' to infer the correct result. This demonstrates the LLM’s ability to extract correlations among inputs to enhance denoising performance. 
\end{itemize}
These observations inspire us to explore the potential of LLMs in CSI reconstruction.

In this paper, we propose a pre-trained LLM-based CSI feedback method. Unlike previous methods that directly learn channel reconstruction from limited CSI datasets, the proposed method uses a pre-trained GPT-2 model to assist with CSI feedback. To mitigate the impact of differences between natural language processing and CSI data processing, we customize the pre-processing, embedding, and post-processing modules for CSI feedback tasks, with only minimal tuning of GPT-2 parameters to minimize training costs. Numerical results validate the effectiveness of applying the pre-trained LLM-based method to CSI feedback tasks with proper modifications.
 
\section{System Model}

We consider a single-cell frequency-division duplexing (FDD) system employing massive MIMO technology.  The BS is equipped with a uniform linear array consisting of $N_{\rm t}$ antennas, where $N_{\rm t} \gg 1$, while the UE  has a single antenna. Orthogonal frequency-division multiplexing (OFDM) is utilized for transmission, with $N_{\rm c}$ subcarriers. The received signal on the $i$-th subcarrier, $y_i \in \mathbb{C}$, can be expressed as: 
\begin{equation} 
y_i={\mathbf{h}}_i^H \mathbf{v}_i x_i + z_i, 
\end{equation}
where $\mathbf{h}_i \in \mathbb{C}^{N_\mathrm{t}\times 1}$  is the downlink channel vector for the $i$-th subcarrier, $\mathbf{v}_i \in \mathbb{C}^{N_\mathrm{t} \times 1}$ is the beamforming vector, $x_i \in \mathbb{C}$is the transmitted symbol, and $z_i \in \mathbb{C}$ represents the additive white Gaussian noise (AWGN). The complete CSI for all $N_\mathrm{c}$ subcarriers is represented by the matrix $\mathbf{H}=[\mathbf{h}_1,\ldots,\mathbf{h}_{N_\mathrm{c}}] \in \mathbb{C}^{N_\mathrm{t}\times N_\mathrm{c}}$. This CSI matrix contains $2N_\mathrm{t}N_\mathrm{c}$ real-valued elements, resulting in substantial communication overhead

To address the significant communication overhead, we adopt a DL-based CSI feedback framework using a one-sided model due to its ease of deployment. As illustrated  in Fig.~\ref{fig:comp}(b), after the UE estimates the CSI matrix $\mathbf{H}$, it employs a random projection matrix to compress the CSI into low-dimensional codewords $\mathbf{s} \in \mathbb{R}^{N_\mathrm{s} \times 1}$:
\begin{equation}
    \mathbf{s}=\mathbf{A}\mathrm{vec}(\mathbf{H}),
\end{equation}
where $\mathbf{A} \in \mathbb{R}^{N_\mathrm{s} \times 2N_\mathrm{t}N_\mathrm{c}}$ is the random projection matrix, and $\mathrm{vec}(\cdot)$ denotes the vectorization of $\mathbf{H}$ after converting it into a real-valued form. The codeword $\mathbf{s}$ is then transmitted to the BS. The compression ratio is defined as 
\begin{equation}
    \gamma=\frac{2N_{\rm t}N_{\rm c}}{N_{\rm s}}.
\end{equation}
Upon receiving the codeword $\mathbf{s}$, the BS first performs a coarse reconstruction of the CSI using the pseudo-inverse of the projection matrix:
\begin{equation}
    \mathbf{H}_\mathrm{in}=\mathrm{devec}(\mathbf{A}^{\dagger}\mathbf{s}).
\end{equation}
where $\mathrm{devec}(\cdot)$ is the inverse operation of $\mathrm{vec}(\cdot)$. To refine the coarsely reconstructed CSI $\mathbf{H}_\mathrm{in}$, a neural network is applied:
\begin{equation}
    \widehat{\mathbf{H}}=f_\mathrm{NN}(\mathbf{H}_\mathrm{in}),
\end{equation}
where $\widehat{\mathbf{H}}$ represents the final reconstructed CSI, and  $f_\mathrm{NN}(\cdot)$ denotes the neural network function used for CSI refinement.

The performance of DL-based CSI feedback models is evaluated using specific loss functions. One widely used metric is the Normalized Mean Square Error (NMSE), defined as:
\begin{equation} 
l_\mathrm{NMSE}{\left(\mathbf{H},\widehat{\mathbf{H}}\right)}=\frac{\|\widehat{\mathbf{H}}-\mathbf{H}\|_{2}^{2}}{\|\mathbf{H}\|_{2}^{2}},
\label{equ:mse} 
\end{equation} 
which measures the reconstruction accuracy of the CSI. Since CSI is crucial for beamforming design, Generalized Cosine Similarity (GCS) is also used to evaluate feedback quality. GCS focuses on the alignment between the true and reconstructed channel vectors and is defined as:
\begin{equation} 
l_\mathrm{CS}{\left(\mathbf{H},\widehat{\mathbf{H}} \right)}=-\frac{1}{N_\mathrm{c}}\sum_{n=1}^{N_\mathrm{c}}\frac{|\widehat{\mathbf{h}}_i^H\mathbf{h}_i|}{\|\widehat{\mathbf{h}}_i \|_2 \|\mathbf{h}_i \|_2},
\label{equ:cs} 
\end{equation} 
where $\widehat{\mathbf{h}}_i$ is the reconstructed channel vector for the $n$-th subcarrier.

\section{LLM for CSI Feedback}
\subsection{Motivation}
Enhancing model capability typically comes at the expense of increased training costs, highlighting a trade-off between performance and computational efficiency. The recent success of pre-trained LLMs across a wide range of downstream tasks demonstrates their remarkable versatility. Trained on data with extremely high diversity, pre-trained LLMs acquire the ability to generalize and solve previously unseen tasks.
As illustrated in Fig.~\ref{fig:comp}(c), the CSI reconstruction task shares structural similarities with the language error correction task. When the CSI matrix is divided into a sequence of CSI vectors for each subcarrier, reconstructing this sequence from noisy observations is analogous to correcting errors in a sequence of words in a sentence. This structural resemblance inspires us to leverage the powerful capabilities of pre-trained LLMs---originally developed for natural language processing---to tackle the CSI reconstruction problem. This approach significantly reduces the training costs associated with building a large, task-specific model for CSI feedback from scratch.

\begin{figure*}[t]
    \centering
    \includegraphics[width=0.95\textwidth]{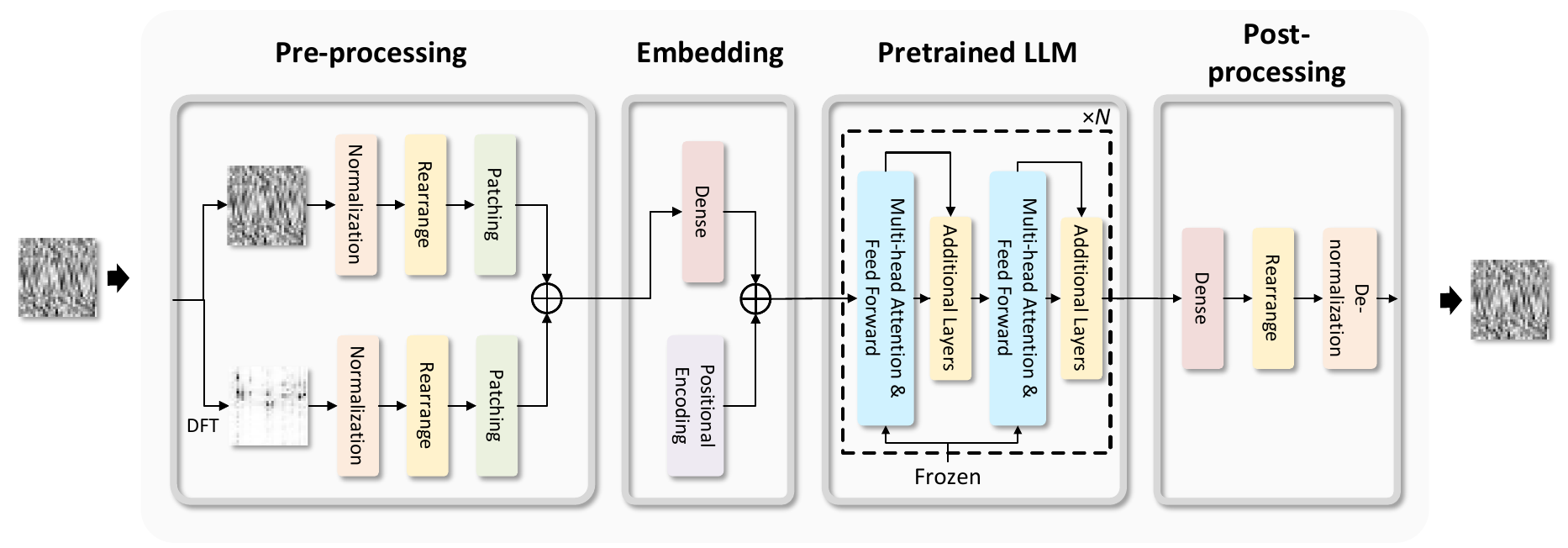} 
    \caption{Detailed architecture of the proposed LLM-based CSI feedback method, consisting of pre-processing, embedding, pre-trained LLM, and post-processing modules.}
    \label{fig:detail}
    \vspace{-0.5cm}
\end{figure*}

\subsection{Key Idea}

We propose a pre-trained LLM-based CSI feedback method. The core idea is to leverage the diverse knowledge embedded in LLMs to achieve high-performance and well-generalized CSI feedback, thereby reducing the need for extensive computational resources and large training datasets. Due to the large number of parameters in LLMs, directly fine-tuning the model on CSI datasets leads to significant computational burdens and may cause overfitting when training data is limited.
To address this, we freeze most of the LLM's parameters to preserve its generalization capability. Since the LLM is pre-trained on text data, it cannot directly process CSI data. Therefore, we design learnable modules to bridge the gap between CSI and textual data representations. The proposed framework, illustrated in Fig.~\ref{fig:detail}, consists of four components: the pre-processing module, the embedding module, the pre-trained LLM, and the post-processing module. Details of each module are provided below.

\subsubsection{Pre-processing} 
To enhance sparsity for more effective compression, the CSI matrix is transformed from the spatial-frequency domain to the angular-frequency domain using the discrete Fourier transform (DFT):
\begin{equation}
    \mathbf{H}_\mathrm{a}= \mathbf{F}_\mathrm{a}{\mathbf{H}_\mathrm{in}},
\end{equation}
where $\mathbf{F}_\mathrm{a} \in \mathbb{C}^{N_{\rm t}\times N_{\rm t}}$ is a DFT matrix. The CSI matrices $\mathbf{H}_\mathrm{in}$ and $\mathbf{H}_\mathrm{a}$ are then normalized to $\widetilde{\mathbf{H}}_\mathrm{in}$ and $\widetilde{\mathbf{H}}_\mathrm{a}$, respectively, to ensure consistency in data range for the embedding module. For notational simplicity, we continue to use $\mathbf{H}_\mathrm{in}$ and $\mathbf{H}_\mathrm{a}$ to represent to represent the normalized versions. Next, each CSI matrix is divided into a sequence of $N_\mathrm{c}$ CCSI vectors in the frequency domain and converted into real-valued forms. Each CSI vector is treated as a ``word'' for the LLM to process:
\begin{equation}
      \mathbf{H}_\mathrm{in}=[{\mathbf{h}}_1,\ldots,{\mathbf{h}}_{N_\mathrm{c}}],  ~~
      \mathbf{H}_\mathrm{a}=[{\mathbf{h}}_\mathrm{a,1},\ldots,{\mathbf{h}}_\mathrm{a,{N_\mathrm{c}}}],
\end{equation}
where ${\mathbf{h}}_i$ and ${\mathbf{h}}_{\mathrm{a},i}$ are in $\mathbb{R}^{2N_\mathrm{t}\times 1}$. To further exploit frequency correlation in the CSI \cite{nie2022time}, a non-overlapping patching operation is applied, generating $\mathbf{H}^\mathrm{p}_\mathrm{in}$ and $\mathbf{H}^\mathrm{p}_\mathrm{a}$. Finally, the angular and spatial domain CSI sequences are combined to form the feature representation for embedding:
\begin{equation}
\mathbf{H}_\mathrm{m}=\mathbf{H}^\mathrm{p}_\mathrm{in}+\mathbf{H}^\mathrm{p}_\mathrm{a},
\end{equation}

\subsubsection{Embedding}
A learnable embedding is used to transform CSI data into token representations suitable for LLM processing. A dense layer maps the input to tokens: 
\begin{equation}
    \mathbf{x}_{\mathrm{m},i}=f_\mathrm{Dense}(\mathbf{h}_{\mathrm{m},i}), 
\end{equation}
where $f_\mathrm{Dense}(\cdot)$ denotes the dense layer and $i=1,\ldots,N_\mathrm{c}$. Since the CSI sequence across different frequencies is ordered, positional encoding is incorporated to capture frequency correlation. The $j$-th element of the positional encoding vector $\mathbf{x}_{\mathrm{pe},i}(j)$ is defined as:
\begin{subequations} 
\begin{align}
\mathbf{x}_{\mathrm{pe},i}(2j)&=\sin( i/10000^{2j/d_\mathrm{em}} ),\\
    \mathbf{x}_{\mathrm{pe},i}(2j+1)&=\cos( i/10000^{2j/d_\mathrm{em}} ),
\end{align}
\end{subequations}
where $d_\mathrm{em}$ is the embedding dimension.
By adding positional encoding to the output of the dense layer, the input to the LLM becomes:
\begin{equation}
\mathbf{x}_{\mathrm{in},i}=\mathbf{x}_{\mathrm{m},i}+\mathbf{x}_{\mathrm{pe},i}
\end{equation}

\subsubsection{Pre-trained LLM}
We employ a pre-trained GPT-2 model for CSI reconstruction \cite{radford2019language}. GPT-2 primarily consists of stacked multi-head attention modules, feed-forward networks, and additional layers, following the typical Transformer architecture \cite{vaswani2017attention}. Pre-trained on the WebText dataset with over 8 million documents, GPT-2 demonstrates strong generalization capabilities across diverse input types.
The pre-trained LLM processes the embedded CSI input as:
\begin{equation}
(\mathbf{x}_{\mathrm{out},1},\ldots,\mathbf{x}_{\mathrm{out},N_\mathrm{c}})=f_\mathrm{LLM}(\mathbf{x}_{\mathrm{in},1},\ldots,\mathbf{x}_{\mathrm{in},N_\mathrm{c}}).
\end{equation}
To preserve the generalization ability from pre-training and minimize computational overhead, most GPT-2 parameters are frozen, except for those in the layer normalization and learnable positional encoding components. 
\subsubsection{Post-processing}
Once the GPT-2 model generates the output tokens, the post-processing module transforms these tokens back into CSI data. Each output token is first mapped to the size of the spatial domain CSI vectors via a dense layer, followed by another dense layer for frequency domain refinement: 
\begin{equation}
(\widehat{\mathbf{h}}_{\mathrm{out},1},\ldots,\widehat{\mathbf{h}}_{\mathrm{out},N_\mathrm{c}})=f_\mathrm{post}(\mathbf{x}_{\mathrm{out},1},\ldots,\mathbf{x}_{\mathrm{out},N_\mathrm{c}}).
\end{equation}
Finally, these vectors are rearranged into the CSI matrix and de-normalized to produce the final reconstructed CSI $\widehat{\mathbf{H}}$.

\section{Simulation Results}
\subsection{Simulation Settings}
\subsubsection{Channel Generation Settings}
We use the QuaDRiGa channel generator to produce CSI samples \cite{jaeckel2014quadriga}. A predefined \texttt{3GPP\_38.901\_UMi\_LOS} scenario is adopted. The base station (BS) is equipped with 32 transmitting antennas operating at 2.655 GHz, corresponding to the ``n7'' frequency band defined in 3GPP TS 38.101-1 \cite{38101}. The system operates with a bandwidth of 70 MHz and 32 subcarriers.
To generate diverse CSI samples, we randomly select 100 distinct circular areas within a cell. The radius of each circular area is 5 meters, and the overall cell radius is 200 meters. For each circular area, a sub-dataset of 10,000 CSI samples is generated. The sub-datasets indexed from 1 to 50 are mixed, shuffled, and split into training, validation, and test datasets in an 8:1:1 ratio. The sub-datasets indexed from 51 to 100 are reserved for evaluating scenario generalization.

\subsubsection{Training Settings}
All neural networks are implemented using PyTorch 2.5.0 and trained on NVIDIA TESLA H100 GPUs. The Adam optimizer is used with a learning rate of $10^{-3}$ and a batch size of 256. The NMSE is employed as the loss function. Both NMSE and GCS are used as performance evaluation metrics.
Each neural network is trained for 200 epochs using the settings described above. The model achieving the best NMSE on the validation set is saved and used for final evaluation. 

For comparison, we introduce a small DNN model, referred to as ``Small Model,'' which shares the same pre-processing, embedding, and post-processing modules as the LLM-based model. In the Small Model, the embedded CSI tokens are concatenated, flattened, and passed through two dense layers, each with 2048 units and LeakyReLU activations. The output is reshaped back to match the original input token size and is processed by the same post-processing module. Additionally, we include the recently proposed TransNet CSI feedback method \cite{cui2022transnet} as a baseline. To further validate the effectiveness of the pre-trained LLM, we introduce a baseline model named ``Identical,'' where the pre-trained LLM module in our proposed framework is replaced by an identical mapping function.

\begin{figure*}[t]
        \centering
	\subfigure []{
		\includegraphics[width=0.48\linewidth]{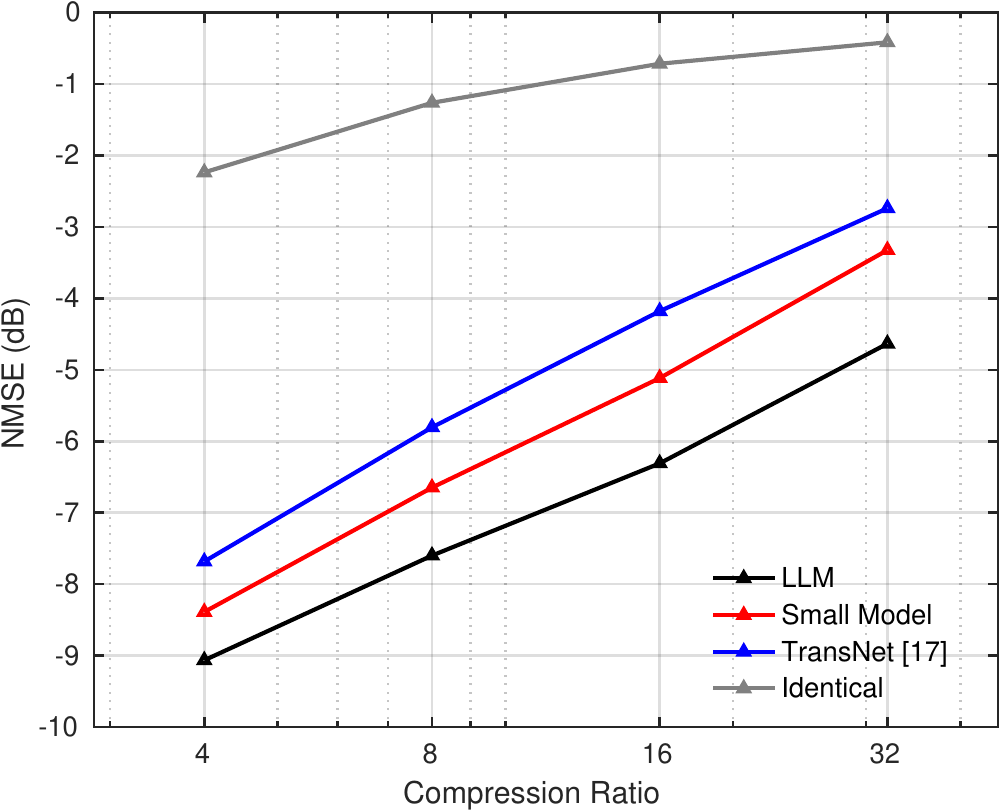}}
	\subfigure []{
		\includegraphics[width=0.49\linewidth]{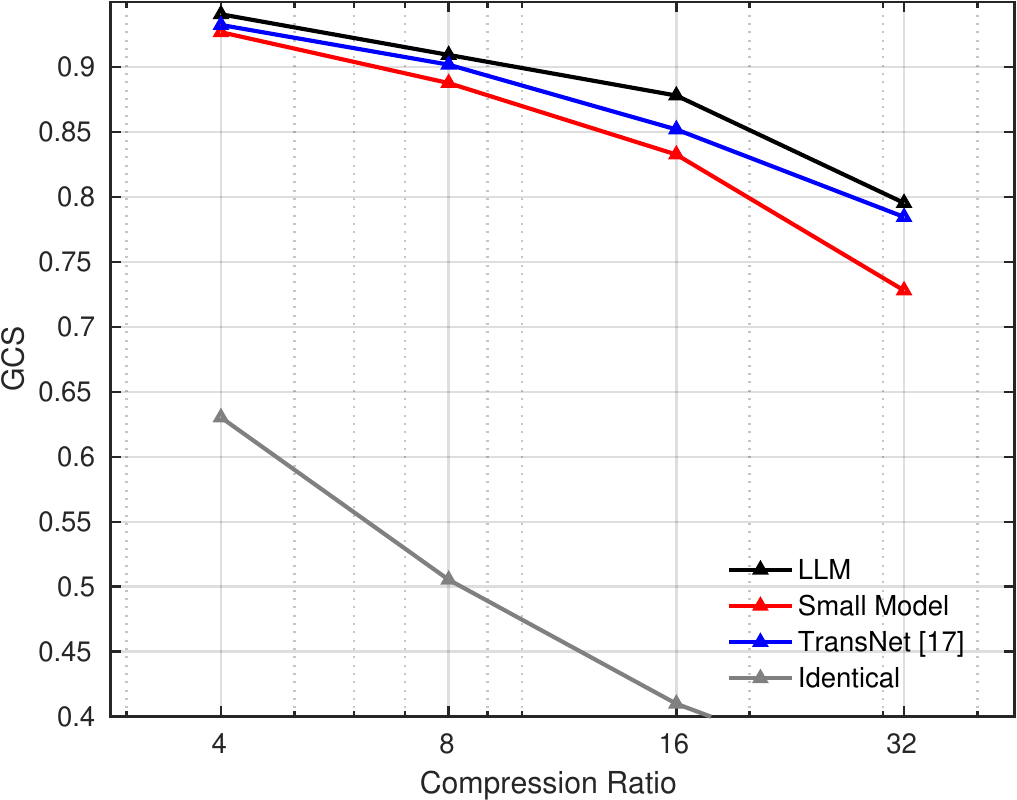}}
        \caption{Comparison of the pre-trained LLM-based CSI feedback method and baseline methods with 50 scenarios and 10,000 samples per scenario.} 
	\label{fig:cr}
\vspace{-0.5cm}
\end{figure*}

\subsection{Performance Improvement}
We evaluate the performance of the pre-trained LLM-based CSI feedback method against the conventional small model-based CSI feedback method using a complex mixed dataset composed of 50 different scenarios. As shown in Fig.~\ref{fig:cr}(a), the proposed method outperforms both the small model-based method and TransNet in terms of NMSE performance. By leveraging the strong capability of the large-scale pre-trained model, which has been trained on abundant and diverse data, the LLM-based model effectively captures complex correlations among CSI elements.
Notably, as the compression ratio increases, the performance gain of the LLM-based method over the small model-based method and TransNet becomes more significant. This is because the small model-based method performs well when the reconstruction task is relatively simple (i.e., at lower compression ratios). However, as the task complexity increases with higher compression ratios, the small model's performance deteriorates sharply due to its limited capacity, whereas the LLM-based method demonstrates greater robustness. This highlights the advantage of the LLM-based method in handling complex CSI reconstruction tasks.
Additionally, the ``Identical'' baseline fails to accurately reconstruct CSI, confirming that the superior performance of the proposed method is primarily attributed to the pre-trained LLM rather than other components. The GCS performance comparison in Fig.~\ref{fig:cr}(b) further confirms that the LLM-based method consistently outperforms the small model-based method and TransNet across various compression ratios.

\begin{figure*}[t]
        \centering
	\subfigure []{
		\includegraphics[width=0.48\linewidth]{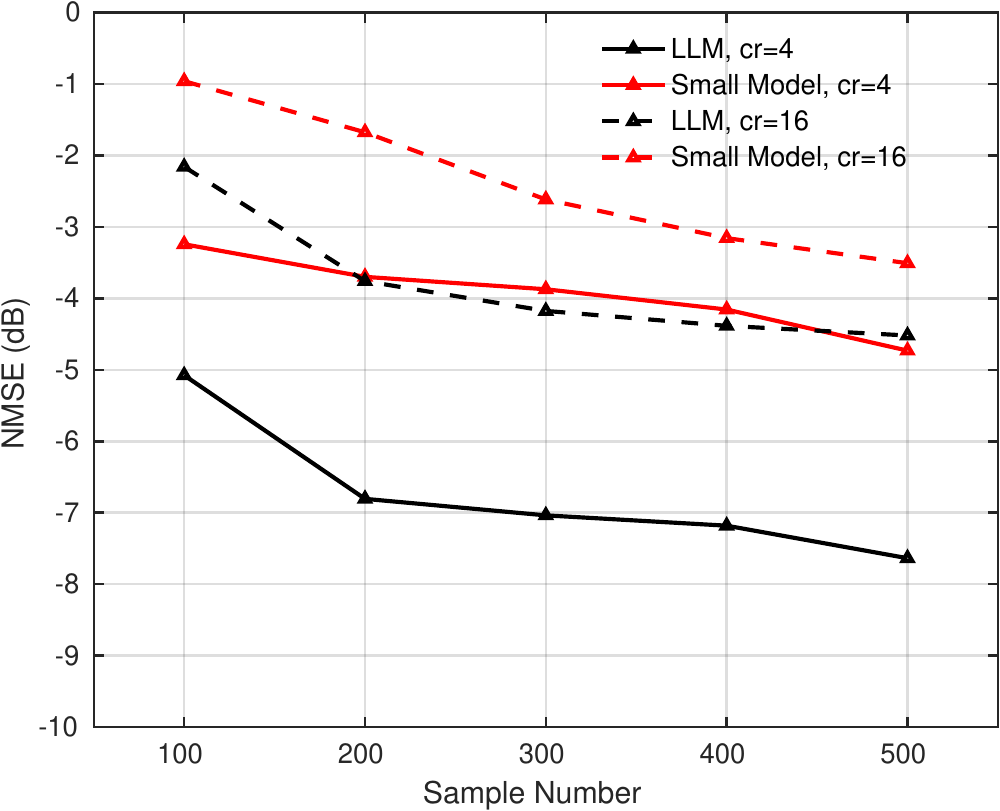}}
	\subfigure []{
		\includegraphics[width=0.49\linewidth]{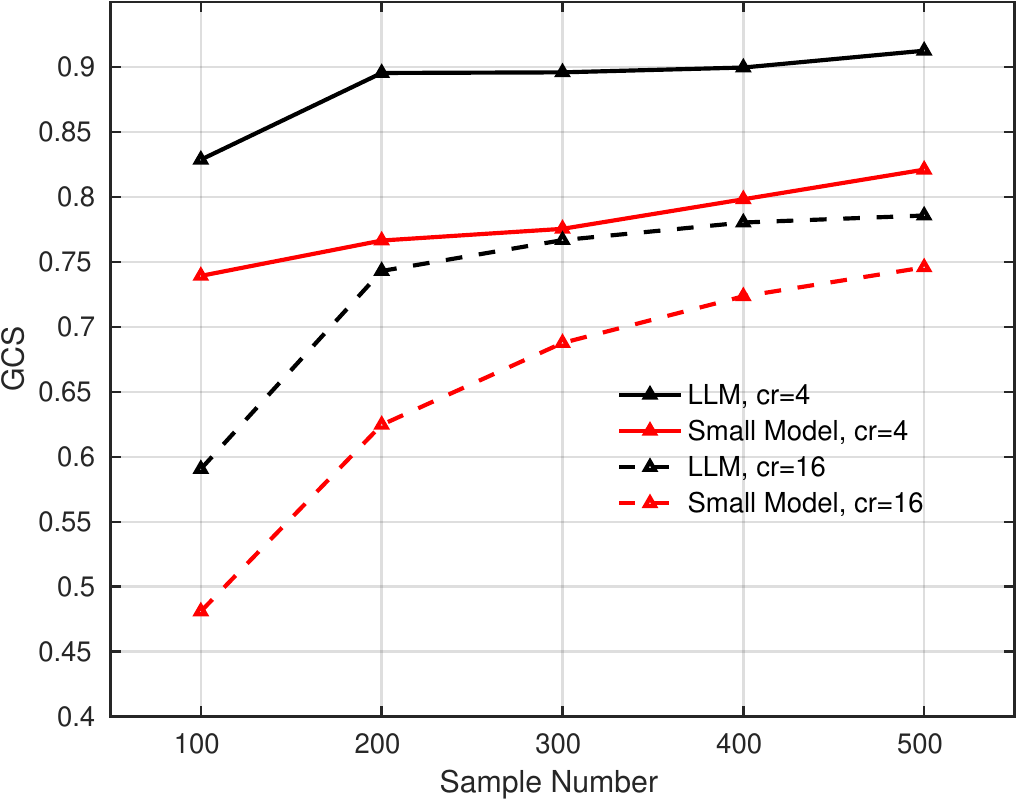}} 
        \caption{Comparison of the pre-trained LLM-based CSI feedback method and baseline methods under limited training data conditions, with 50 scenarios used for evaluation.}
	\label{fig:sample}
\vspace{-0.5cm}
\end{figure*}

\subsection{Training Cost Reduction}
We further examine the impact of the pre-trained LLM on reducing training costs. Due to the high cost of CSI data collection and the significant computational resources required for training on large-scale datasets, we compare the performance of the pre-trained LLM-based method with the small model-based method under limited training data conditions.
As shown in Fig.~\ref{fig:sample}(a), the performance of the small model-based method degrades significantly when trained on insufficient data. For example, when only 100 to 500 samples per scenario are used for training with a compression ratio of 4, the NMSE performance drops from 5.15 dB to 3.66 dB compared to training on the full dataset. In contrast, the pre-trained LLM-based method maintains strong performance even with limited data. Specifically, when trained with 500 samples per scenario, the LLM-based method experiences only a 0.75 dB performance drop compared to full dataset training.
A similar trend is observed for GCS performance. When more than 200 CSI samples per scenario are used, the LLM-based method achieves performance comparable to full dataset training. The minimal performance gap between full and small dataset training demonstrates that the language knowledge learned from text data can be effectively transferred to assist in CSI reconstruction, significantly reducing training costs with only minor performance degradation.

\begin{figure*}[t]
        \centering
	\subfigure []{
		\includegraphics[width=0.48\linewidth]{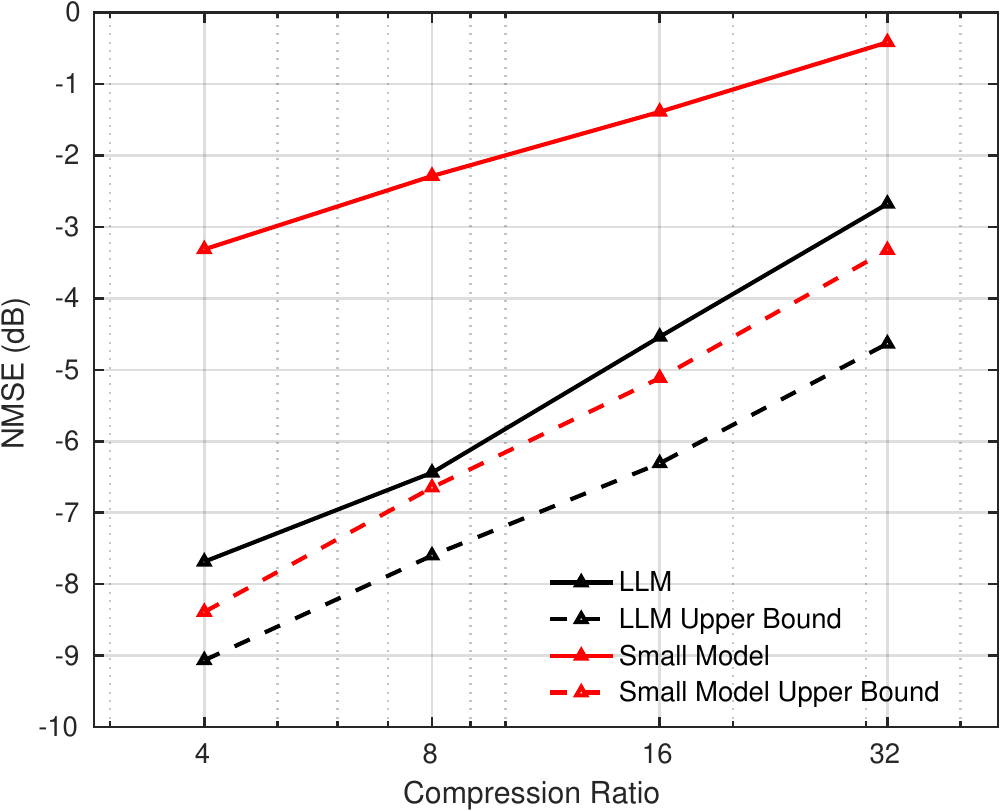}}
	\subfigure []{
		\includegraphics[width=0.49\linewidth]{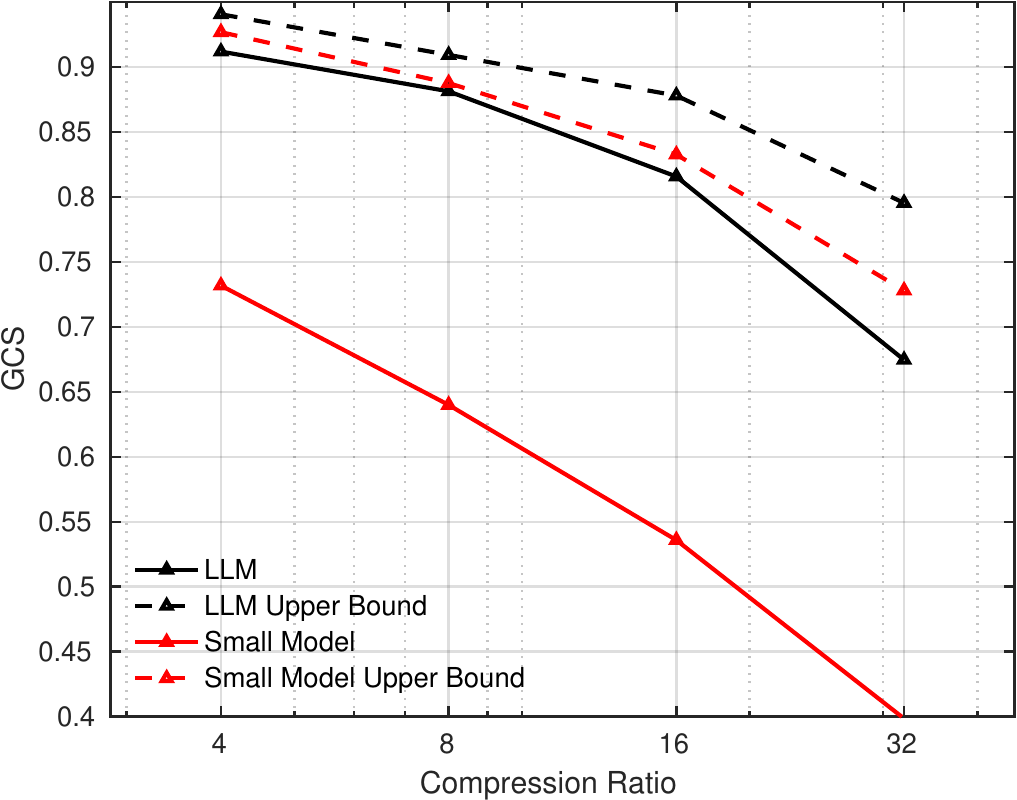}}
        \caption{Comparison of the generalization capability between the pre-trained LLM-based CSI feedback method and baseline methods. The models are trained on the mixed dataset from scenarios indexed 1 to 50 and tested on the mixed dataset from scenarios indexed 51 to 100, with 10,000 samples per scenario.} 
	\label{fig:gen}
\vspace{-0.5cm}
\end{figure*}

\subsection{Generalization Improvement}
Finally, we investigate the generalization capability of different CSI feedback methods. Models are trained using the mixed datasets from scenarios indexed 1 to 50 and tested on the datasets from scenarios indexed 51 to 100. For comparison, upper bounds are established by training and testing on the same datasets from scenarios indexed 51 to 100.
As shown in Fig.~\ref{fig:gen}, the proposed pre-trained LLM-based method generalizes well to unseen scenarios, with only a small performance gap compared to the upper bound. In contrast, the small model exhibits significant performance degradation relative to its upper bound. When evaluated using GCS, the performance gap between the LLM-based method and the upper bound becomes even smaller, further highlighting the LLM-based method's strong generalization capability.
Due to its limited capacity and the diversity of the training data, the small model tends to converge to simpler solutions that overfit the training scenarios and struggle to adapt to new ones. In contrast, the pre-trained LLM equips the proposed method with the ability to approximate a more generalized solution for CSI reconstruction, resulting in superior generalization across different scenarios.

\section{Conclusion}
In this paper, we proposed a pre-trained LLM-based CSI feedback method. To bridge the gap between natural language processing and CSI data, we designed specialized pre-processing, embedding, and post-processing modules to adapt the LLM for CSI feedback tasks.
Simulation results demonstrate that the proposed pre-trained LLM-based method achieves high-performance CSI feedback even with limited training data. This confirms that the knowledge learned from natural language processing can be effectively transferred to the CSI feedback task, offering a promising direction for intelligent wireless communication systems.

\bibliographystyle{IEEEtran}
\bibliography{refer1}

\begin{thebibliography}{10}
\providecommand{\url}[1]{#1}
\csname url@samestyle\endcsname
\providecommand{\newblock}{\relax}
\providecommand{\bibinfo}[2]{#2}
\providecommand{\BIBentrySTDinterwordspacing}{\spaceskip=0pt\relax}
\providecommand{\BIBentryALTinterwordstretchfactor}{4}
\providecommand{\BIBentryALTinterwordspacing}{\spaceskip=\fontdimen2\font plus
\BIBentryALTinterwordstretchfactor\fontdimen3\font minus \fontdimen4\font\relax}
\providecommand{\BIBforeignlanguage}[2]{{%
\expandafter\ifx\csname l@#1\endcsname\relax
\typeout{** WARNING: IEEEtran.bst: No hyphenation pattern has been}%
\typeout{** loaded for the language `#1'. Using the pattern for}%
\typeout{** the default language instead.}%
\else
\language=\csname l@#1\endcsname
\fi
#2}}
\providecommand{\BIBdecl}{\relax}
\BIBdecl

\bibitem{wang2024tutorial}
Z.~Wang, J.~Zhang, H.~Du, D.~Niyato, S.~Cui, B.~Ai, M.~Debbah, K.~B. Letaief, and H.~V. Poor, ``{A Tutorial on Extremely Large-Scale MIMO for 6G: Fundamentals, Signal Processing, and Applications},'' \emph{IEEE Communications Surveys \& Tutorials}, vol.~26, no.~3, pp. 1560--1605, 2024.

\bibitem{marzetta2015massive}
T.~L. Marzetta, ``{Massive MIMO: An Introduction},'' \emph{Bell Labs Technical Journal}, vol.~20, pp. 11--22, 2015.

\bibitem{lu2014overview}
L.~Lu, G.~Y. Li, A.~L. Swindlehurst, A.~Ashikhmin, and R.~Zhang, ``{An Overview of Massive MIMO: Benefits and Challenges},'' \emph{IEEE Journal of Selected Topics in Signal Processing}, vol.~8, no.~5, pp. 742--758, 2014.

\bibitem{love2008overview}
D.~J. Love, R.~W. Heath, V.~K. N.~Lau, D.~Gesbert, B.~D. Rao, and M.~Andrews, ``{An overview of limited feedback in wireless communication systems},'' \emph{IEEE Journal on Selected Areas in Communications}, vol.~26, no.~8, pp. 1341--1365, 2008.

\bibitem{metzler2016denoising}
C.~A. Metzler, A.~Maleki, and R.~G. Baraniuk, ``{From Denoising to Compressed Sensing},'' \emph{IEEE Transactions on Information Theory}, vol.~62, no.~9, pp. 5117--5144, 2016.

\bibitem{wen2018deep}
C.-K. Wen, W.-T. Shih, and S.~Jin, ``{Deep Learning for Massive MIMO CSI Feedback},'' \emph{IEEE Wireless Communications Letters}, vol.~7, no.~5, pp. 748--751, 2018.

\bibitem{guo2022overview}
J.~Guo, C.-K. Wen, S.~Jin, and G.~Y. Li, ``{Overview of Deep Learning-Based CSI Feedback in Massive MIMO Systems},'' \emph{IEEE Transactions on Communications}, vol.~70, no.~12, pp. 8017--8045, 2022.

\bibitem{yu2020ds}
X.~Yu, X.~Li, H.~Wu, and Y.~Bai, ``{DS-NLCsiNet: Exploiting Non-Local Neural Networks for Massive MIMO CSI Feedback},'' \emph{IEEE Communications Letters}, vol.~24, no.~12, pp. 2790--2794, 2020.

\bibitem{wang2021compressive}
J.~Wang, G.~Gui, T.~Ohtsuki, B.~Adebisi, H.~Gacanin, and H.~Sari, ``{Compressive Sampled CSI Feedback Method Based on Deep Learning for FDD Massive MIMO Systems},'' \emph{IEEE Transactions on Communications}, vol.~69, no.~9, pp. 5873--5885, 2021.

\bibitem{peng2024deep}
Z.~Peng, Z.~Li, R.~Liu, C.~Pan, F.~Yuan, and J.~Wang, ``Deep learning-based csi feedback for ris-aided massive mimo systems with time correlation,'' \emph{IEEE Wireless Communications Letters}, vol.~13, no.~8, pp. 2060--2064, 2024.

\bibitem{radford2019language}
A.~Radford, J.~Wu, R.~Child, D.~Luan, D.~Amodei, I.~Sutskever \emph{et~al.}, ``{Language Models are Unsupervised Multitask Learners},'' \emph{OpenAI blog}, vol.~1, no.~8, p.~9, 2019.

\bibitem{nie2022time}
\BIBentryALTinterwordspacing
Y.~Nie, N.~H. Nguyen, P.~Sinthong, and J.~Kalagnanam, ``{A time series is worth 64 words: Long-term forecasting with transformers},'' \emph{arXiv:2211.14730}, 2022. [Online]. Available: \url{http://arxiv.org/abs/2211.14730}
\BIBentrySTDinterwordspacing

\bibitem{liu2024llm4cp}
B.~Liu, X.~Liu, S.~Gao, X.~Cheng, and L.~Yang, ``{LLM4CP: Adapting Large Language Models for Channel Prediction},'' \emph{Journal of Communications and Information Networks}, vol.~9, no.~2, pp. 113--125, 2024.

\bibitem{vaswani2017attention}
A.~Vaswani, N.~Shazeer, N.~Parmar, J.~Uszkoreit, L.~Jones, A.~N. Gomez, L.~Kaiser, and I.~Polosukhin, ``{Attention Is All You Need},'' \emph{Advances in Neural Information Processing Systems}, 2017.

\bibitem{jaeckel2014quadriga}
S.~Jaeckel, L.~Raschkowski, K.~B{\"o}rner, and L.~Thiele, ``{QuaDRiGa: A 3-D multi-cell Channel Model with Time Evolution for Enabling Virtual Field Trials},'' \emph{IEEE Trans. Antennas Propag.}, vol.~62, no.~6, pp. 3242--3256, Jun. 2014.

\bibitem{38101}
\BIBentryALTinterwordspacing
{3GPP TS 38.101-1}, ``{NR}; user equipment ({UE}) radio transmission and reception; {Part} 1: {Range} 1 {Standalone} ({Release} 17),'' Tech. Rep., Mar. 2021. [Online]. Available: \url{https://portal.3gpp.org/desktopmodules/Specifications/SpecificationDetails.aspx?specificationId=3283}
\BIBentrySTDinterwordspacing

\bibitem{cui2022transnet}
Y.~Cui, A.~Guo, and C.~Song, ``{TransNet: Full Attention Network for CSI Feedback in FDD Massive MIMO System},'' \emph{IEEE Wireless Communications Letters}, vol.~11, no.~5, pp. 903--907, 2022.

\end{thebibliography}
\end{document}